# Reciprocity approach for calculating the Purcell effect for emission into an open optical system

K. M. SCHULZ[1,*], D. JALAS[1], A. Y. PETROV[1,2] AND M. EICH[1,3]

[1]*Institute of Optical and Electronic Materials, Hamburg University of Technology, Eissendorfer Strasse 38, 21073 Hamburg, Germany.*
[2] *ITMO University, 49 Kronverkskii Ave., 197101, St. Petersburg, Russia*
[3] *Institute of Materials Research, Helmholtz-Zentrum Geesthacht, Max-Planck-Strasse 1, Geesthacht, D-21502, Germany*
*\*marvin.schulz@tuhh.de*

**Abstract:** Based on the reciprocity theorem, we present a formalism to calculate the power emitted by a dipole source into a particular propagating mode leaving an open optical system. The open system is completely arbitrary and the approach can be used in analytical calculations but may also be combined with numerical electromagnetic solvers to describe the emission of light sources into complex systems. We exemplify the use of the formalism in numerical simulations by analyzing the emission of a dipole that is placed inside a cavity with connected single mode exit waveguide. Additionally, we show at the example of a practical ring resonator system how the approach can be applied to systems that offer multiple electromagnetic energy decay channels. As a consequence of its inherent simplicity and broad applicability, the approach may serve as a powerful and practical tool for engineering light-matter-interaction in a variety of active optical systems.

## 1. Introduction

In 1946, E. M. Purcell predicted that the spontaneous emission rate of a light source is not solely an intrinsic property of the source but is affected by the optical environment [1] – an effect that is now referred to as Purcell effect. The Purcell factor, defined as

$$F_p = \frac{P_{system}}{P_0}, \qquad (1)$$

where $P_{system}$ denotes the power of an emitter radiated into a particular optical system and $P_0$ the power of the same emitter radiated into vacuum free space, is a common figure of merit to describe the emission enhancement induced by feedback of the source with a particular optical system. For spontaneous emission, the Purcell factor can alternatively be defined as $F_p = \frac{\tau_0}{\tau_{system}}$ where $\tau_0$ is the spontaneous emission lifetime in vacuum and $\tau_{system}$ the lifetime of the emitter in the particular system of interest.

Today, the Purcell effect is the corner stone for a variety of applications including single-photon sources [2], integrated quantum optics [3,4], nanoscale lasers [5], active metamaterials [6], biotechnological devices for enhanced fluorescence intensity [7], ultra-fast modulated LEDs [8] and nonlinear wave mixing devices [9].

In cavities, the emission enhancement is approximately given by $F_p \sim \frac{Q}{V}$ where Q is the quality factor and V is the mode volume that is often approximated by the cavity volume [1]. Plasmonic resonators display the highest Purcell factors reported, in the order of $10^3$, that are derived from a small $V$ rather than high Q [10–13]. Emission enhancement is not limited to narrowband resonant systems. Waveguides, for instance, provide a broadband enhancement effect due to local field enhancement in the guided optical mode that may exist over a large frequency range. [5,14–16].

The interaction between the emitter and its environment is formally described by Fermi's golden rule which states that the probability for spontaneous emission is proportional to the (photonic) local density of states (LDOS) [17,18]. For any electromagnetic environment, the LDOS is rigorously given by the imaginary part of the electric field Green dyadic at the position of the emitter [19,20]. To obtain the electric field Green dyadic, an exact solution of the wave equation with a point source in the considered optical system has to be calculated. The system should be open to allow for power extraction. In order to probe different positions inside the system new Green's function calculations are required [10,21]. In addition, the Green's function may have an evanescent part that is real valued and diverging at the emitter position and does not lead to power emission. In order to resolve this evanescent part, a fine discretization close to the emitter is required in the numerical simulations [10,21]. To avoid the Green's function calculation, alternative approaches appeared to identify the local electric fields at the emitter position. For example, the LDOS can be identified as the number of eigenmodes per frequency interval in closed systems. In this definition each mode is normalized by its energy and the contribution of each eigenmode to the sum is weighted by the square of the local electric mode field at the emitter location [21]. However, for open systems, such as a cavity leaking radiation into the outside world, the concept of normal eigenmodes and their electric field normalization is not well defined. Formally, for such systems one obtains a diverging normalization integral since the eigenmode fields expand to infinity in space outside of the system due to their leakage [22]. In practice, this fundamental problem is often ignored by choosing a finite integration volume. This may indeed yield accurate results for high Q-factor systems weakly coupled to the outside world, however, for strongly coupled (low-Q factor) systems, the results may become inaccurate [22]. One way to account for this problem is to introduce the notion of quasi-normal modes with complex frequencies and refine the mode volume definition allowing for complex valued mode volumes [22–24]. While these approaches well describe open resonant systems they cannot be applied to waveguides or other arbitrary non-resonant open systems.

It has recently been shown in the context of single molecule detections that the power emitted from a molecule into a single mode fiber can be elegantly calculated using the reciprocity theorem of electromagnetic theory [25]. Here, we show how a reciprocity approach can be used to calculate the emission enhancement for emitters coupled to arbitrary resonant or non-resonant open optical systems. In our approach, we excite the optical system from outside and determine the ratio between the inserted power and the local electric field. In doing so, we can calculate the work this excitation does on a dipole at its position. Using the Lorentz reciprocity relation we can identify the power emitted by this dipole into the outside world. Our formalism is based on the reciprocity theorem and can be used to describe emission of light sources into any lossy and non-lossy system for that the electromagnetic energy leaving the system can be described by propagating modes. These systems also include open resonant systems that are coupled to propagating modes, without limitation on the coupling strength. We exemplify the formalism with an example of a leaky cavity emitting radiation into a single mode waveguide and for ring resonator case that represents a system with multiple electromagnetic energy decay channels. For the these systems, our ansatz only requires the excitation of the system with the well-defined waveguide mode of interest. It does not require a solution for the localized eigenmodes of the resonator. Consequently, the notion of quasi normal modes or mode-volume definitions for open systems and Q factor calculations are not required. The approach can be used analytically or in electromagnetic solvers to analyze the emission into complex open systems.

## 2. Reciprocity approach

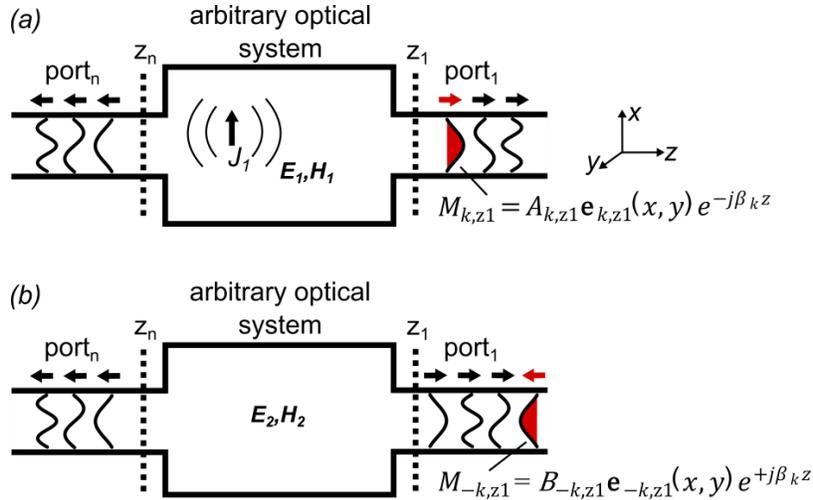

Fig. 1 (a) an oscillating current density distribution $J_1$ emits radiation into an arbitrary optical system resulting in a field distribution given by $E_1, H_1$. In the shown case, the excitation energy can leave the system through the exit $\text{port}_1$ that has a cross section $z_1$ and other ports summarized by one port denoted as $\text{port}_n$ with a cross section $z_n$. At the cross section $z_1$ the fields $E_1$ and $H_1$ are expanded into an orthogonal set of modes. The power leaving the system through the cross section $z_1$ that is carried by a particular mode $M_k$ shall be calculated. (b) back excitation of the mode of interest $M_k$ into the system with arbitrary amplitude $B_{-k}$. The reciprocal mode has an opposite propagation direction and is denoted as $M_{-k}$. Excitation of the mode results in a field distribution in the system given by $E_2, H_2$. The reciprocity approach enables us to calculate the power of a source emitted into mode $M_k$ from the reciprocally excited field distribution $E_2, H_2$.

Figure 1(a) shows a current density distribution $J_1$ that emits radiation into an arbitrary optical system with two attached exit ports denoted as $port_1$ and $port_n$ that have a cross section $z_1$ and $z_n$, respectively. In the steady state, $J_1$ excites an electromagnetic field distribution in the system given by $E_1, H_1$. The fields $E_1, H_1$ are Green's functions for the electric and magnetic wave equations for the system (i.e. solution to the inhomogeneous wave equation including a source term) and consider effects such as back reflections of emitted fields from the system boundaries or exit ports. We impose no further assumption on the source except that its intrinsic properties are maintained by the interaction with the environment, that is, emission frequency and current are not affected by the interaction and quantum electrodynamic effects such as Lamb shift and Rabi splitting do not occur [26,27]. We describe the fields leaving the system via the exit ports as propagating modes. In principle, the derivation can be made for the system with a single exit port. In Fig. 1(a) we added $port_n$ to represent other energy decay channels in the system and to demonstrate that this add-on does not affect the derivation as any other additional port would not change it either. Thus, we expand the outgoing fields $E_1$ and $H_1$, excited by $J_1$, at the exit cross sections $z_1$ and $z_n$ into the normal sets of modes given by

$$\begin{pmatrix} E_1(x,y,z_1) \\ H_1(x,y,z_1) \end{pmatrix} = \sum_i A_{i,z1} \begin{pmatrix} e_{i,z1}(x,y)e^{-j\beta_i z_1} \\ h_{i,z1}(x,y)e^{-j\beta_i z_1} \end{pmatrix} \quad (2)$$

at cross section $z_1$ and

$$\begin{pmatrix} E_1(x,y,z_n) \\ H_1(x,y,z_n) \end{pmatrix} = \sum_i A_{-i,zn} \begin{pmatrix} e_{-i,zn}(x,y)e^{+j\beta_i z_n} \\ h_{-i,zn}(x,y)e^{+j\beta_i z_n} \end{pmatrix} \quad (3)$$

at the cross section $z_n$ where $A_i$ and $A_{-i}$ are complex amplitudes carrying the phase information. The fields are normalized such that $|A_{i/-i}|^2$ is the power carried in a particular mode $i/-i$. The sign in the index i indicates the propagation direction of the modes in z direction at cross section $z_1$ and $-z$ direction at cross section $z_n$, respectively.

In the following, we calculate the power that the polarization source $J_1$ emits into a particular mode of the orthogonal set at the cross section $z_1$. We denote this mode with index $i = k$ as $M_{k,z1} = A_{k,z1} e_{k,z1}(x,y)e^{-j\beta_k z}$. For this purpose, we consider the reciprocal excitation of the mode $M_{k,z1}$ with arbitrary amplitude $B_{-k}$ shown as case 2 in Fig. 1(b). We denote the reciprocal mode as $M_{-k,z1} = B_{-k,z1} e_{-k,z1}(x,y)e^{+j\beta_k z}$ since it has an opposite propagation direction compared to the emitted mode $M_{k,z1}$. Excitation of $M_{-k,z1}$ in case 2 will result in a field distribution in the system that is given by the field vectors $E_2, H_2$, representing a solution to the homogenous (i.e. source free) wave equation for the system. In the reciprocal excitation case (Fig. 2(b)), the fields $E_2, H_2$ can be expanded into one incoming mode, that is the excitation mode $M_{-k,z1}$, and several outgoing modes at the cross section $z_1$ and other cross section $z_n$ given by

$$\begin{pmatrix} E_2(x,y,z_1) \\ H_2(x,y,z_1) \end{pmatrix} = B_{-k,z1} \begin{pmatrix} e_{-k,z1}(x,y)e^{+j\beta_k z_1} \\ h_{-k,z1}(x,y)e^{+j\beta_k z_1} \end{pmatrix} + \sum_i B_{i,z1} \begin{pmatrix} e_{i,z1}(x,y)e^{-j\beta_i z_1} \\ h_{i,z1}(x,y)e^{-j\beta_i z_1} \end{pmatrix} \quad (5)$$

and

$$\begin{pmatrix} E_2(x,y,z_n) \\ H_2(x,y,z_n) \end{pmatrix} = \sum_i B_{-i,zn} \begin{pmatrix} e_{-i,zn}(x,y)e^{+j\beta_i z_n} \\ h_{-i,zn}(x,y)e^{+j\beta_i z_n} \end{pmatrix}, \quad (6)$$

respectively. We use the reciprocity theorem to relate the fields for case 1 and 2 [28]

$$\oiint E_1 \times H_2 - E_2 \times H_1 dA = \int_V E_2 J_1 \, dV. \tag{7}$$

We write the left hand side of Eq. (7) as

$$\int_{z=z_1}(E_1 \times H_2 - E_2 \times H_1)n_z dxdy - \int_{z=z_n}(E_1 \times H_2 - E_2 \times H_1)n_z dxdy = \int_V E_2 J_1 \, dV \tag{8}$$

in order to evaluate the surface integrals at the cross sections $z_1$ and $z_n$ on the left hand side of the equation. Plugging in the fields and using orthogonality of the modes as well as

$$e_i = e^*_{-i} \text{ and } h_i = h^*_{-i} \tag{9}$$

for a lossless material at the cross section were modes are defined [29], the only non-zero term is the product between the mode that we send in case 2, $M_{-k,z1}$, and the outgoing mode in case 1, $M_{k,z1}$,. We note that so far no assumption on loss was made and a general lossy system was considered. The presented approach well includes the effect of material loss in the full system, only at the position in the port where the modes are expanded into an orthogonal set a loss-free condition is assumed (Eq. (9)).

Hence, we arrive at

$$-2A_{k,z1}B_{-k,z1}\int_{z=z_1}\left(e(x,y)_{k,z1} \times h(x,y)^*_{k,z1}\right)n_z dxdy = \int_V E_2 J_1 \, dV, \tag{10}$$

where the contribution from the port $z_n$ is vanished. Taking into account the applied power normalization of the propagating modes this is simplified to

$$-4A_{k,z1}B_{-k,z1} = \int_V E_2 J_1 \, dV, \tag{11}$$

where the integral on the left hand side of Eq. (10) has a numerical value of 2 according to the power normalization. For an excitation with the reciprocal mode $M_{-k,z1}$ in case 2, the square of the local electric field $|E_2(r)|^2$ will at all positions $r$ be proportional to the power carried by $M_{-k,z1}$ into the port $z_1$ $P_{-k,z1} = |B_{-k,z1}|^2$. Thus, without loss of generality, we can write

$$E_2(r) = \alpha_k(r)B_{-k,z1}, \tag{12}$$

where $\alpha_k(r)$ is a complex proportionality constant relating the local field $E_2(r)$ and the mode amplitude $B_{-k,z1}$ at the cross section $z_1$. $\alpha_k(r)$ also includes the phase difference between the local electric field $E_2(r)$ and the mode amplitude $B_{-k,z1}$ at the cross section $z_1$. Plugging this into the right hand side of Eq. (11) we obtain:

$$-4A_{k,z1}B_{-k,z1} = \int_V B_{-k,z1}\alpha_k(r) \cdot J_1 \, dV. \tag{13}$$

Using $|A_{k,z1}|^2 = P_{k,z1}$ which is the power emitted into mode $k$ we write

$$P_{k,z1} = \frac{1}{16}\left|\int_V \alpha_k(r) \cdot J_1 \, dV\right|^2. \tag{14}$$

We now assume the current density distribution $J_1$ is a single discrete dipole placed at position $r_0$. In this case we can write $\omega \alpha_k(r_0)p = \int_V \alpha_k(r) \cdot J_1 dV$ [30] where $p$ is the dipole moment and $\omega$ is the angular frequency. Thus, we arrive at

$$P_{k,z1} = \frac{1}{16}\omega^2 |\alpha_k(r_0) \cdot p|^2. \tag{15}$$

This is a general solution for the emitted power into mode $k$. We note that its evaluation requires only the knowledge of the field distribution in the optical system excited by the reciprocal mode and the intrinsic dipole moment of the source.

If the electric field polarization coincides with the orientation of the dipole, we can evaluate the scalar product as $\boldsymbol{\alpha}_k(\boldsymbol{r_0}) \cdot \boldsymbol{p} = \alpha_k(\boldsymbol{r_0})p$. Hence, for a dipole placed at position $\boldsymbol{r_0}$ the power emitted into $M_k$ is given by

$$P_{k,z1} = \frac{1}{16}\omega^2 p^2 |\alpha_k(\boldsymbol{r_0})|^2. \tag{16}$$

We define the Purcell factor $F_k(\boldsymbol{r_0})$ for emission into mode $k$ by

$$F_k(\boldsymbol{r_0}) = \frac{P(\boldsymbol{r_0})_{k,z1}}{P_0} = \frac{3\pi c}{4\mu_0 \omega^2}|\alpha_k(\boldsymbol{r_0})|^2, \tag{17}$$

where $c$ is the speed of light, $\mu_0$ is the vacuum permeability and

$$P_0 = \frac{\mu_0}{12\pi c}\omega^4 p^2, \tag{18}$$

is the power of the dipole emitted into homogenous free space according to Larmor's formula [30]. Note that the Purcell factor defined here only measures the power emitted into a single mode of interest, $M_{k,z1}$, leaving the optical system with respect to free space emission. The total Purcell factor may be higher and is given by the sum of powers emitted into all electromagnetic energy decay channels of the system that could be represented by other modes but also ohmic losses [31].

We take a closer look at the expression derived in Eq. (15). We note that according to Eq. (15), the emitted power of the dipole into $M_{k,z1}$ depends only on the emission frequency $\omega$, the dipole moment $\boldsymbol{p}$ and the proportionality constant $\boldsymbol{\alpha}_k(\boldsymbol{r_0})$. While $p$ and $\omega$ are intrinsic properties of the emitter, all information about the effect of the external optical environment on the emitted power is condensed in $\boldsymbol{\alpha}_k(\boldsymbol{r_0})$. For emission of a dipole placed into a cavity coupled to a port, that is the effect of mode volume and Q-factor. For emission of a dipole placed inside a waveguide [32], $\boldsymbol{\alpha}_k(\boldsymbol{r_0})$ contains the effect of slow light, reflecting interfaces and degree of light confinement. Consequently, we call $\boldsymbol{\alpha}_k(\boldsymbol{r_0})$ the mode interaction parameter since it measures the strength of light-matter interaction for emission into a particular propagating mode. $\boldsymbol{\alpha}_k(\boldsymbol{r_0})$ can be determined according to Eq. (12) by relating the local field amplitude of the electric field to the amplitude of the reciprocal mode. This relation may be found analytically or by employing numerical electromagnetic solvers. Thus, emission into propagating modes leaving arbitrary open systems can be analyzed.

Note that the derivation is made here for a dipole source. Dissipation of energy into the system due to higher order multipole moments of the source [33] can be described within the formalism if the current density distribution $\boldsymbol{J}_1$ in Eq. (14) is chosen accordingly. The formalism can also describe distributed emitters radiating into the same propagating mode. For the case of distributed polarization sources given by $\boldsymbol{\varrho}(\boldsymbol{r})$ and dipole moment $\boldsymbol{p}$ given by $\boldsymbol{p} = \int_V \boldsymbol{\varrho}(\boldsymbol{r}) \, dV$, Eq. (14) is expressed as

$$P_{k,z1} = \frac{1}{16}\left|\int_V \boldsymbol{\alpha}(\boldsymbol{r}) \cdot \boldsymbol{\varrho}(\boldsymbol{r}) \, dV\right|^2, \tag{19}$$

where $\boldsymbol{\varrho}(\boldsymbol{r})$ also carries the phase information of the coherent polarization distributed in space. If the distributed sources oscillate incoherently, as for example in case of thermal excitation, then just the powers emitted by individual dipoles (Eq. (14)) are added up separately.

## 3. Emission of a dipole into a cavity with attached exit-waveguide

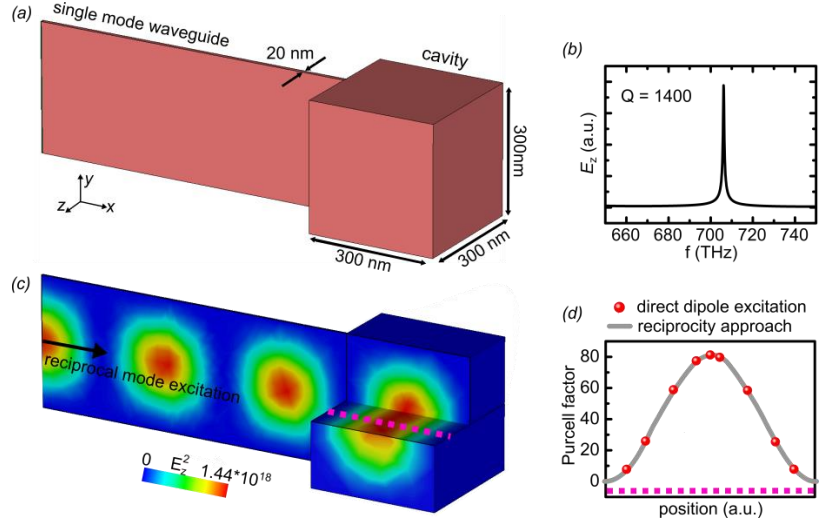

Fig. 2 (a) investigated example system. A narrow single- mode waveguide is attached to a cubic cavity. Both, waveguide and cavity have walls consisting of perfect electric conductor (i.e. loss less metal). (b) Spectrum of the cavity coupled to the waveguide. The cavity Q-factor is 1400 and the resonance frequency is 707 THz. (c) Electric field distribution $E_z^2(r)$ after excitation of the system with the waveguide mode. The waveguide mode is launched into the system from the left hand side as indicated by the arrow. (d) Purcell factor for a dipole placed at different positions inside the cavity along an exemplary cutting line indicated by the dashed purple line in (c). Solid line: Purcell factor as obtained from the reciprocity approach. Red dots: Purcell factor as obtained from emission of a modelled dipole placed at discrete positions along the cutting line.

We exemplify the use of the formalism in numerical simulations by exploring the emitted power of a dipole placed inside a leaky cavity as shown in Fig. 2(a). The cavity has a side length $d = 300$ nm and walls consisting of perfect electric conductor (i.e. lossless metal). A hollow rectangular waveguide with a width of $w = 20$ nm and height of $h = 300$ nm is attached to the cavity and represents the only exit for electromagnetic energy leaving the system. The waveguide only supports a single mode at the emission frequencies under study. Figure 2(b) shows the cavity spectrum for emission of an emitter placed inside the cavity. The cavities' resonance frequency is $f_0 = \frac{c}{\sqrt{2}*d} = 707$ THz and a Q-factor $Q = \frac{f_0}{\Delta f} = 1400$ where $\Delta f$ denotes the full width at half maximum of the resonance peak.

In the following the power emitted into the propagating waveguide mode by a dipole that is placed inside the cavity shall be calculated using the reciprocity approach. Since in the chosen example the single mode waveguide is the only decay channel provided in the system, we expect that the power coupled through the waveguide mode is equal to the total power emitted by the dipole. If the waveguide supported more modes, the total emitted power of the dipole would be given by the sum of powers emitted into all individual waveguide modes. We choose a dipole oriented in z-direction and consider emission at the cavity resonance frequency $f_0$. We remark, however, that the approach does not distinguish between resonant or off-resonant emission and describes these cases equally exact. Numerical simulations are carried out using the CST MICROWAVE STUDIO package. We excite the system with the reciprocal waveguide mode and determine the square of the mode interaction parameter

$\alpha^2(\mathbf{r})$ inside the cavity according to Eq. (12), by relating the local electric field amplitude $E_z^2(\mathbf{r})$ to the time-average power used for excitation of the mode (in our case 0.5W). The electric field distribution $E_z^2(\mathbf{r})$ for the reciprocal mode excitation is shown in Fig. 2(c). We observe a standing wave pattern in the system since the excitation fields and the fields back reflected from the system, that travel in opposite direction, superimpose. After determining $\alpha^2(\mathbf{r})$, Eq. (16) and Eq. (17) are used to calculate the position dependent emitted power and Purcell factor of the dipole, respectively. In Fig. 2(d) we show the Purcell factor inside the cavity along an exemplary cutting line as indicated in Fig. 1(c) (dashed purple line).

In order to demonstrate the validity of the reciprocity approach, we compare our results to the case where a model dipole orientated in z-direction is placed at discrete positions along the cutting line and its total emitted power is measured directly. To measure the emitted power of the dipole, we use the induced impedance numerical method [34,35]. Alternatively, for loss less systems, the emitted power of the source can be determined by integration of the Poynting vector across the port exit area [11,36,37]. We implement the dipole as a subwavelength metal antenna and the effect of the environment on its radiation is measured in terms of the antenna radiation resistance. This method measures the total emitted power of the dipole into all electromagnetic decay channels offered by the system [6,34]. In our case the total emitted power by the dipole should be equal to that emitted into the waveguide mode, since the waveguide mode is the only decay channel. In order to calculate the Purcell factor (Eq. (1)), the power of the same dipole emitted into free space vacuum is measured and used for normalization. The results are shown in Fig. 2(d). As can be seen, the results from the direct dipole excitation and the reciprocity approach agree within the accuracy of the numerical simulation for all positions under study confirming the applicability of the reciprocity approach. We emphasize that, as opposed to the direct dipole excitation method, the reciprocity approach enables us to map out the emitted power into the waveguide mode for the whole system based on a single simulation and the emitter does not have to be explicitly modelled as for example in [10]. The latter step is usually computationally expensive and prone to errors as the implemented emitter is much smaller than the optical wavelength and therefore sensitive to the discretization error. In addition, for multimode systems, the power dissipated into different modes cannot be distinguished when measuring the induced impedance of the model dipole. In addition, for multimode systems, the power dissipated into different modes cannot be distinguished when measuring the induced impedance of the model dipole.

## 4. Engineering emission enhancement in integrated quantum light sources

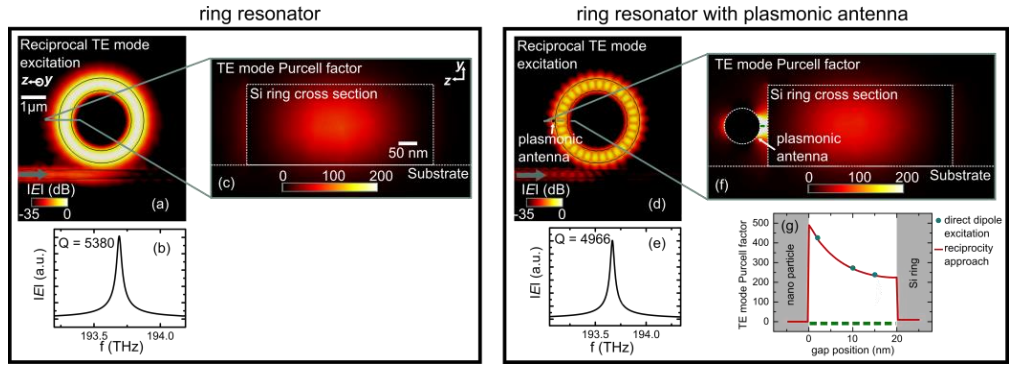

Fig. 3 (a) integrated dielectric ring resonator coupled to an adjacent straight dielectric waveguide with geometric parameters defined in the text. The shown field distribution in the ring and the waveguide originates from excitation of the TE mode in the straight waveguide. (b) Electric field spectrum and Q-factor of the resonator. (c) TE mode Purcell factor evaluated for the cross section of the ring indicated in (a). The TE mode Purcell factor shows the power of an emitter at any position of the system that is radiated into the propagating TE mode of the straight waveguide. It is normalized to the power of the same emitter radiated into vacuum free space. The calculation is based on the reciprocal mode excitation shown in (a) and Eq. 17. (d) Field distribution from excitation of the TE-mode for dielectric ring resonator with adjacent plasmonic nanoantenna. (e) Electric field spectrum and Q factor of the resonator with adjacent plasmonic nanoantenna. (f) TE mode Purcell factor evaluated at the cross section indicated in (d). (g) TE mode Purcell factor along the line cut at the hot spot between the waveguide and the nanoparticle indicated in (f) (green dashed line).

As a second example we now explore emission of a light source into a practical system that offers multiple electromagnetic energy decay channels. The example considers emission of an integrated light source into an on-chip integrated photonic circuit [38]. We show now that the reciprocity approach provides a reverse way for identifying radiative enhancement hot spots for emission into particular waveguide modes of the system. The system is composed of an on-chip integrated ring resonator coupled to a straight waveguide, as shown in Fig. 3(a). A similar system was recently investigated experimentally in the framework of quantum information processing [39]. An emitter, e.g. a single photon source, placed in vicinity of the ring, couples electromagnetic energy into the localized resonator mode which is then, in turn, coupled to a propagating waveguide mode. The propagating mode distributes the emitter energy within the photonic circuit. In our example we employ the standard silicon-on-insulator platform at telecom wavelength of 1.5 µm [40]. The ring resonator and waveguide have a height of 220 nm. The waveguide has a width of 350 nm and supports a TE propagating mode and a TM propagating mode. The ring resonator has a width of 500 nm and outer radius of 1715 nm. The distance between the waveguide and the ring at the position with the smallest separation is 280 nm which is the optimum coupling distance for critical coupling. The ring resonator and the waveguide are coated with a polymer with refractive index of 1.6 that can host emitting molecules.

An emitter placed close to the ring resonator can dissipate its energy into several radiative energy decay channels of the system. First of all, the emitter can directly emit to free space modes or into the substrate. At the same time, it can couple to the ring resonator and then to the propagating TE mode of the adjacent waveguide.

We now calculate the power of the emitter radiated into the propagating TE mode of the straight waveguide. For this purpose, we reciprocally excite the waveguide mode of interest in the straight waveguide. The obtained field distribution at resonance frequency of the ring is shown in Fig. 3(a). From the electric field spectrum we determine a Q-factor of 5380 as shown in Fig. 3(b). Based on the reciprocally excited field distribution, we apply Eq. (16) to calculate the radiated power into the propagating TE mode for an emitter oriented in z-direction placed at any position in the system. Figure 3(c) shows the result for a certain cross section of the ring waveguide. We plot the TE mode Purcell factor (Eq.17) which normalizes the power emitted into the TE mode of the waveguide to the power of the same emitter radiated into vacuum free space.

We note that the TE mode Purcell factor reaches values of up to 100 for positions inside and at the left hand side of the waveguide. Thus, in order to increase the power emitted into the propagating TE mode of the straight waveguide the emitter should ideally be placed inside the ring resonator or on the left hand side of the ring. Generally, hot spots in the reciprocally obtained false - color maps might originate from resonances, confinement effects, field discontinuities at material boundaries, interference and dispersion effects (slow light) and complex coupling interplay in coupled systems. It should be noted that the TE mode Purcell factor displayed here only measures the power of the source emitted into the TE mode of the straight waveguide. The total dissipated power of the source might be higher and is given by the sum of powers emitted into all available decay channels of the system including other radiation modes and energy of the source dissipated into ohmic loss [31] (quenching). The power emitted into the TM mode of the waveguide could also be calculated employing the same method (reciprocal excitation of the TM mode in the system).

We now illustrate how multiple coupled systems and plasmonic systems can also be well described by the approach. For this purpose, we place an additional plasmonic antenna close to the ring resonator. Such plasmonic- dielectric hybrid systems have recently been explored in the literature in conjunction with radiative emission enhancement [41]. Figure 3 (d) shows the field distribution for the same resonator system where an additional spherical gold nano particle with a radius of 48 nm is placed close to the ring resonator. The simulation considers loss of the metallic component according to optical parameters taken from [42]. The reciprocally excited electromagnetic field distribution now reveals a standing wave pattern in the ring due to scattering of the gold particle which leads to forward and backward travelling waves in the ring. The scattering outside the ring, due to the presence of the particle, slightly reduces the Q-factor of the ring resonator to Q = 4966 as shown in Fig. 3(e). However, despite the reduction in Q-factor, the adjacent plasmonic particle greatly enhances the power of the emitter coupled to the TE mode of the straight waveguide for certain emitter positions in the system. The TE mode Purcell factor map in Fig. 3(f) reveals an emission hot spot in the gap between the nanoparticle and the ring resonator. In this gap, the TE mode Purcell factor exceeds values of 200. In Fig. 3(g) we provide a detailed evaluation of the enhancement in the gap. In order to confirm our results, we place a dipolar current source with a length of 1 nm at certain positions in the gap between the particle and the ring and directly calculate its power emitted into the propagating TE mode of the straight waveguide. To distinguish the power of the source emitted into the TE mode from the power emitted into other modes (TM mode, free space modes) we calculate the field overlap of the emitted fields with the TE mode of the waveguide at the simulation boundary. The power results are referenced to the power of the same source emitted into vacuum free space. As can be seen, the results from this direct dipole excitation method and reciprocity approach agree within the numerical accuracy, well confirming the applicability of the reciprocity approach to describe emission into particular modes leaving arbitrarily complex, strongly or weakly coupled lossy systems that feature multiple decay channels.

We also discuss the application of other methods to the presented examples. The direct dipole excitation method (equivalent to a Green's function approach) can only be used to calculate the emission at discrete positions in the system. Several simulations are required to map out the decay in the full system and the simulation time scales linearly with the number of positions probed as opposed to the reciprocity approach. We find that for the explored system with optimized discretization to calculate the emission for a single position in the discrete dipole approach takes a time comparable to a reciprocal mode excitation. The reciprocity approach is therefore computationally more efficient when spatial mapping of the emission enhancement in complex systems is required and especially when position tolerances for emission enhancement have to be identified. As opposed to the direct dipole excitation, the reciprocal mode excitation can additionally exploit system symmetries to further reduce the simulation time. We also note that obtained false - color maps provide a much deeper and more intuitive insight into the underlying enhancement processes, that can be used for light-matter engineering, compared to individually probed points.

We remark that coupled resonator systems can alternatively be treated by a quasi-normal mode approach with complex valued resonance frequencies and a complex valued mode volume definition [43,44]. For coupled resonator systems as explored in [43] and [44] we estimate the same computational effort when using the reciprocity approach. In our examples, we show however that this degree of conceptual complexity is not required if one wants to calculate the power emitted into a particular mode leaving the open optical system. Quasi-normal modes, complex valued cavity mode volumes and resonant frequencies and Q-factors do not have to be defined. In addition, the reciprocity approach can also be used to describe emission from non-resonant structures into particular modes leaving the system and the power emitted into particular modes can be well distinguished. In fact, the same approach can be used to describe emission into modes from completely arbitrary structures.

## 5. Conclusion

In conclusion we presented a formalism to calculate the power emitted by a light source into particular modes leaving an open optical system. The formalism is based on the Lorentz reciprocity theorem and is applicable to any system for which the electromagnetic energy leaving the system can be described by discrete propagating modes. The formalism can describe emission from arbitrary lossy, resonant and non-resonant structures into modes of an exit channel. Well described exit channels can for instance be represented by: RF waveguide modes, plasmonic waveguide and dielectric waveguide modes, modes of single and multimode fibers but also discrete freely propagating modes such as Gaussian beams or plane waves that serve as approximate solution for certain systems. In the approach, the optical system is excited from outside with the mode of interest and the work this excitation does on a dipole placed inside the system is calculated. Using the Lorentz reciprocity relation we can identify the power 3emitted by this dipole into the mode of interest. For resonant cavities leaking radiation into propagating modes, the same formalism does not distinguish between weakly or strongly coupled resonators and resonant or off-resonant emission and is exact in describing these cases. It also does not require mode volume definitions for open resonator modes or Q-factor calculations or mode overlap calculations. We exemplified the use of the formalism in numerical simulations in a semi-analytical approach by analyzing emission into a coupled cavity system and showed that the formalism is a powerful and practical tool to characterize emission into propagating modes leaving arbitrarily complex open systems. We also showed that the approach can be applied to analyze emission into lossy systems with multiple energy decay channels. For such systems the power emitted into particular modes can be analyzed and well distinguished. Comparison with the direct dipole excitation evidences the applicability of the approach also for lossy systems. As a consequence of its simplicity and broad applicability, the formalism may serve as a useful tool to engineer light-matter interaction in a variety of active optical devices.

## 6. Acknowledgement

We gratefully acknowledge CST, Darmstadt, Germany for support with their MICROWAVE STUDIO software.

## 7. Funding

We gratefully acknowledge funding by the Deutsche Forschungsgemeinschaft (DFG, German Research Foundation) - Projektnummer 392323616 and the Hamburg University of Technology (TUHH) in the funding programme "Open Access Publishing".